# Heat Capacity of Isotopically Enriched $^{28}$Si, $^{29}$Si and $^{30}$Si in the Temperature Range 4K <$T$< 100K

A. Gibin, G.G. Devyatykh, A.V. Gusev

*Institute of Chemistry of High-Purity Substances, Russian Academy of Sciences,*

*603950, Nizhny Novgorod, Russia*

R.K. Kremer, M. Cardona,

*Max-Planck-Institut für Festkörperforschung, D-70569 Stuttgart, Germany*

H.-J. Pohl

*VITCON Projectconsult GmbH, 07745 Jena, Germany*

The heat capacity of isotopically enriched $^{28}$Si, $^{29}$Si, $^{30}$Si samples has been measured in the temperature range between 4K and 100K. The heat capacity of Si increases with isotopic mass. The values of the Debye temperature for the three isotopic varieties of silicon have been determined. Good agreement with the theoretical dependence of the Debye temperature on isotopic mass has been found.

## 1. Introduction

The investigation of the effect of the isotopic composition on the physical properties of a large number of semiconductors and insulators has been the subject of vivid research since high-purity single crystals with isotopic compositions different from the natural one became available during the past 15 years. Due to their technological importance, the semiconductors Si and Ge have occupied a special place in these investigations. Recently, for example, the thermal conductivity of isotopically enriched Si and Ge has attracted particular attention because of its potential applications [1-6]. The heat capacity of isotopically enriched Ge samples has been reported in [7]. In ref. [8] *ab initio* calculations of the isotopic dependence of the heat capacity of Si, Ge and diamond are reported.

The heat capacity of diamond with several isotopic compositions has been recently measured [9].

In ref. [10] the first heat capacity measurements of isotopically enriched silicon $^{28}$Si with an isotopic purity of 99.88% were reported for the temperature range 1-140 K. The values of the heat capacity, obtained for $^{28}$Si and for $^{nat}$Si using 2 – 6mg samples significantly deviate from those available in the literature for $^{nat}$Si [11, 12], casting serious doubts on the reliability of those measurements. To the best of our knowledge, the heat capacity of single crystals of the other stable isotopes of silicon ($^{29}$Si and $^{30}$Si), has not been reported.

In this paper we present measurements of the heat capacity of highly isotopically enriched samples of $^{28}$Si, $^{29}$Si and $^{30}$Si and compare these results with the *ab initio* calculations of Sanati *et al.* [8].

| Sample | Isotopic composition | Average mass (amu) |
|---|---|---|
| $^{28}$Si | 99.983% $^{28}$Si + 0.014% $^{29}$Si +0.003% $^{30}$Si | 27.977 |
| $^{nat}$Si | 92.23% $^{28}$Si + 4.67% $^{29}$Si +3.10% $^{30}$Si | 28.086 |
| $^{29}$Si | 97.58% $^{29}$Si + 2.15% $^{28}$Si + 0.27% $^{30}$Si | 28.958 |
| $^{30}$Si | 98.68% $^{30}$Si + 0.62% $^{29}$Si + 0.70% $^{28}$Si | 29.954 |

Table I The isotopic compositions and average isotopic masses of the investigated samples

## 2. Experimental

The measurements were carried out on single crystals whose isotopic composition is given in Table I. At the MPI für FKF in Stuttgart the heat capacities were determined for samples of mass 0.1 – 0.3 g with a home-built Nernst calorimeter (using Lake Shore Cernox CX1050 thermometers [13]) and for smaller samples (20 – 50 mg) with a commercial PPMS (Physical Property Measurement System, Quantum Design, 6325 Lusk Boulevard, San Diego) employing the relaxation method [14]. The measurement errors did not exceed 3% in the temperature range $4 < T < 10K$ and 1% for $T > 20K$. The data obtained with the different setups in the 20-200K range agree with each other within the experimental error.

At the ICHPS of the Russian Academy of Sciences the measurements were carried out in an adiabatic calorimeter. In order to measure the temperature, a carbon resistance thermometer (Lake Shore CGR-1-1000 [13]) was used. The error in the determination of the heat capacity did not exceed 0.4 % throughout the whole available temperature range (20-75K). For a direct comparison with the $^{28}$Si data, the heat capacity of a crystal of high-purity $^{nat}$Si of the same mass and similar geometry was also measured. The measurements at the ICHPS of the RAS were carried out on a sample of $^{28}$Si with a mass of 59.2g manufactured from the same crystal used for the measurements at the MPI für FKF.

We note that the $^{28}$Si samples were of a higher chemical purity than the $^{29}$Si and $^{30}$Si samples. The content of gas-forming impurities (O, C) in the $^{28}$Si sample was $2 \cdot 10^{16}$ at./cm$^3$ and that of electrically active impurities less than $1 \cdot 10^{14}$ at./cm$^3$. The electrically active impurities were determined by measuring the Hall effect and by photothermal ionization spectroscopy. The content of metal impurities ($<10^{-5}$ at.-%) was obtained by laser mass-spectroscopy.

## 3. Results and Discussion

**Temperature Region 4 -10K**

A pronounced effect of the isotopic composition on the heat capacity is observed in the low-temperature regime. For high-purity $^{nat}$Si the following dependence of heat capacity upon temperature applies in the temperature range from 2 to 9 K [12]:

$$C_P = \frac{12\pi^4 R}{5}\left(\frac{T}{\Theta_D}\right)^3, \qquad (1)$$

where $\Theta_D$ is the Debye temperature. For $^{nat}$Si the $\Theta_D$ relevant to Eq. (1) has been measured to be 636K [15]. Assuming that the force constants do not dependent on the isotopic mass, $\Theta_D$ for an average isotopic mass $M_i$ is related to $M_i$ by:

$$\Theta_D^i \propto \frac{1}{\sqrt{M_i}} \qquad (2)$$

and the ratio of two heat capacities $C_{1,2}$ corresponding to the isotopic masses $M_{1,2}$ at low temperatures is determined by the relation:

$$\frac{C_1}{C_2} = \left(\frac{M_1}{M_2}\right)^{3/2} \qquad (3)$$

The experimental data in the temperature range 4 to 10 K are presented as a $C_P/T - T^2$ plot in Fig. 1. In this temperature range the data are very well described by the expression

$$C_P = \gamma T + \beta T^3 \qquad (4)$$

with the term $\beta T^3$ representing the lattice heat capacity and $\gamma T$ an additional contribution most likely due to the presence of free carriers. The heat capacities of $^{28}$Si are very close to the data for $^{nat}$Si which agree well with the data of Flubacher *et al.*[11]. As in the case of high-purity $^{nat}$Si, the heat capacity of our $^{28}$Si is determined by the lattice contribution $\beta T^3$ only. A linear contribution to the heat capacity is observed for $^{29}$Si and $^{30}$Si which is small for $^{29}$Si but quite substantial in the case of $^{30}$Si. It can be ascribed to the heat capacity of free-carriers at a concentration level of $6\times10^{19}$ at.cm$^{-3}$ [15]. According to thermopower measurements this sample is p-type. The coefficients $\beta$ and $\gamma$ and the Debye temperature $\Theta_D$ obtained from the least squares fits are compiled in Table II.

| isotope | $\beta$ ($10^{-6}$ J/mol K$^4$) | $\Theta_D$ (0) (K) | $\gamma$ ($10^{-5}$ J/mol K$^2$) |
|---|---|---|---|
| $^{28}$Si | 7.35 ± 0.04 | 641.9 ± 2 | — |
| $^{29}$Si | 7.81 ± 0.03 | 629.0 ± 1 | 0.15 ± 0.1 |
| $^{30}$Si | 8.29 ± 0.04 | 616.6 ± 2 | 6.5 ± 0.2 |
| $^{nat}$Si | 7.44 ± 0.04 | 639.4 ± 1 | — |

Table II Coefficients $\beta$ and $\gamma$ obtained from a fit of Eq. (4) to the experimental heat capacities between 4 and 10K. We have also listed the Debye temperatures $\Theta_D$ (0) derived from $\beta$ using Eq. (1).

The ratios of the Debye temperatures $\Theta_D$ (0) for the three Si isotopes calculated from the lattice contributions $\beta T^3$ to the heat capacities are in reasonable agreement with the predictions of Eq. (2):

$$\Theta_D(^{28}\text{Si}) / \Theta_D(^{29}\text{Si}) = 1.021; \quad \sqrt{M_{Si29} / M_{Si28}} = 1.018$$
$$\Theta_D(^{29}\text{Si}) / \Theta_D(^{30}\text{Si}) = 1.041; \quad \sqrt{M_{Si29} / M_{Si28}} = 1.035 \tag{5}$$

The $\Theta_D$ measured for $^{nat}$Si (639.4 ± 1 K) agrees with that reported by Keesom and Seidel (636 K) [13].

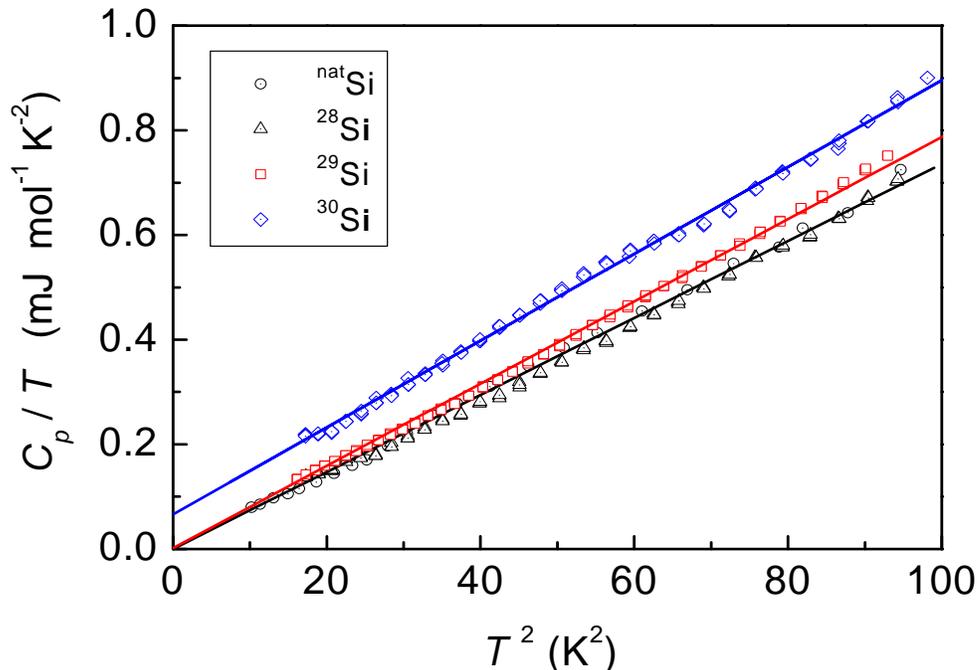

Fig. 1 Heat capacity of natural and isotopically enriched Si in the temperature range 4K < $T$ < 10K. We have used the standard plot of $C_P/T$ vs $T^2$, which allows the separation of the vibrational from the electronic contributions.

**Temperature Region 10-80K**

We show in Fig. 2 our results for $C_p(T)$ normalized to $T^3$ in order to facilitate a comparison with the Debye theory which predicts a temperature independent $C_p/T^3$ at low temperatures. This figure displays our data for $^{nat}$Si and $^{28}$Si in comparison with what we believe to be the most reliable measurements for $^{nat}$Si in the literature and the results of the *ab initio* calculations by Sanati *et al.* [8]. Our results for $^{nat}$Si agree within scatter with the data of Flubacher *et al.* [11]. Below 35K and above 50K the heat capacities of $^{nat}$Si and $^{28}$Si agree within the experimental error. Between 35K and 50K the heat capacities of $^{28}$Si are lower than those of $^{nat}$Si. The difference amounts to ~1.5% in the temperature range 38 – 44 K, somewhat larger than 0.5%, expected from an estimate according to Eq. (3).

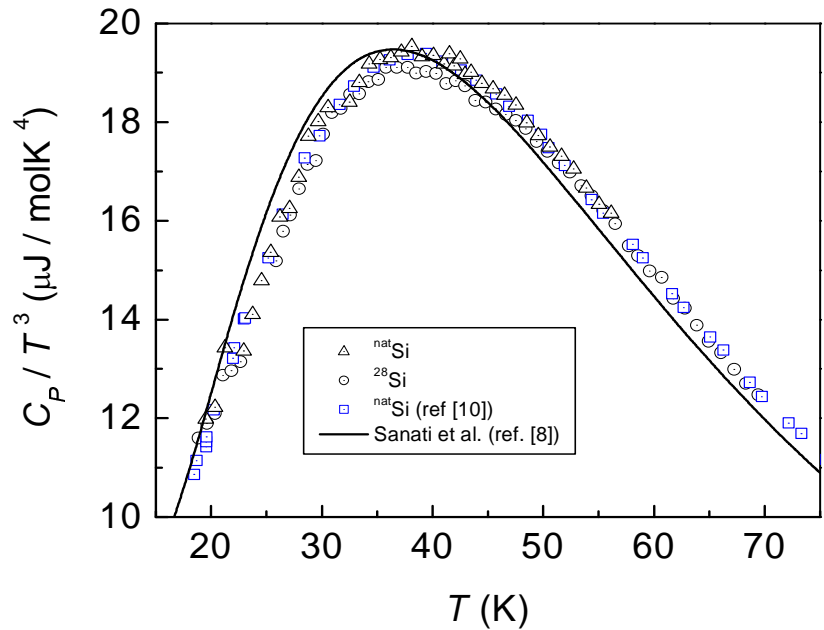

Fig. 2. Heat capacity of $^{nat}$Si and $^{28}$Si compared to the literature data of Flubacher *et al.* [11]. The solid line represents the results of the ab-initio calculations by Sanati *et al.* [8].

Fig. 3 shows the measured heat capacities of all stable Si isotopes compared with those of $^{nat}$Si and the results of the *ab initio* calculations by Sanati *et al.*[8]. The isotope effect is most pronounced around the maximum which results from the strong singularities found for the TA phonons at the edge of the Brillouin zone [16]. The maxima shift from 38.6(3)K to 38.1(3)K and 37.3(4)K for $^{nat,28}$Si, $^{29}$Si, and $^{30}$Si, respectively. The heat capacity maxima correspond to minima in the effective Debye temperatures $\Theta_D(T)$ which occur at 463 K, 456 K, and 447 K for $^{28}$Si, $^{29}$Si and $^{30}$Si, respectively. The measured maximum values of $Cp/T^3$ show good agreement with the *ab initio* calculations (except for $^{29}$Si) while the temperature at which the maxima occur in the calculations are shifted by about 6% to lower temperatures with respect to the measured points.

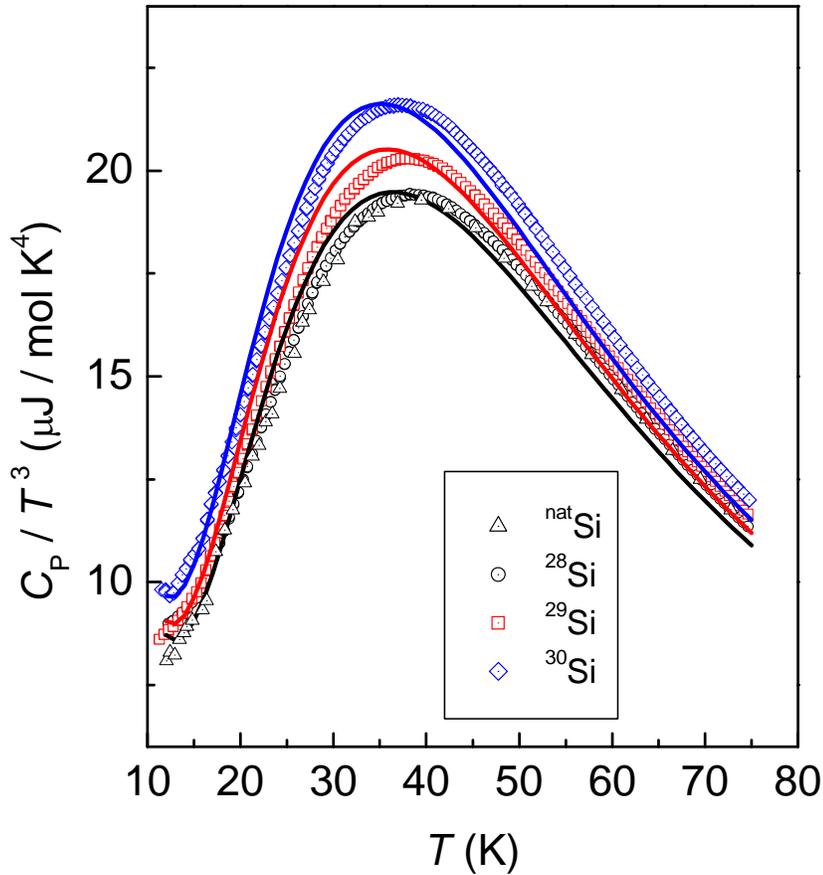

Fig.3. Heat capacity of natural and isotopically enriched Si The solid lines represent the calculations given by Sanati *et al.* [8]

The logarithmic derivative with respect to the isotopic mass was shown in ref. [8] to be related to the logarithmic derivative with respect to the temperature by the equation:

$$\frac{d\ln(C_p/T^3)}{d\ln M} = \frac{1}{2}\left(3 + \frac{d\ln(C_p/T^3)}{d\ln T}\right) \qquad (6)$$

Eq. 6 and the temperature dependence of the heat capacity of $^{nat}$Si were used to calculate $d\ln(C_p/T^3)/d\ln M$. Good agreement with the theoretical results (see Fig. 6 in Ref. [8]) was obtained.

Our measured heat capacities of Si with different isotopic masses enables us to calculate directly the logarithmic derivative with respect to the isotope mass (LHS of Eq. (6)). A similar analysis was recently carried out for diamonds with different isotope compositions [9]. Our results for Si are compared in Fig. 4 with the *ab initio* calculated temperature dependence of $d\ln(C_P/T^3)/d\ln M$.

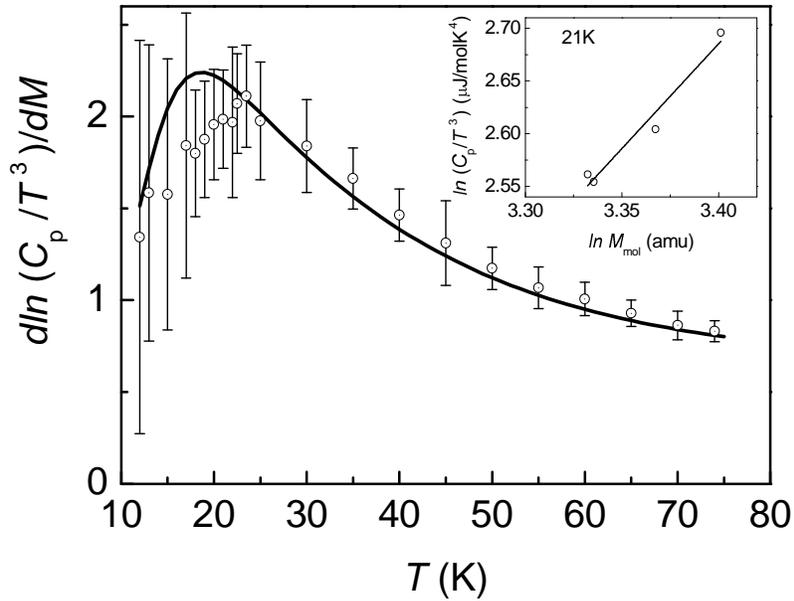

Fig. 4. Logarithmic derivative of $C_p/T^3$ with respect to the isotopic mass of Si. The circles were obtained from the experimental data using a procedure illustrated in the inset. The solid line represents the temperature dependence calculated according to [8].

The agreement of the directly measured and calculated logarithmic derivatives is excellent for $T>20K$. It appears that the maximum in the experimental data is shifted to higher temperatures and the decrease shows up also at somewhat higher temperatures than theoretically predicted, although the large error bars for $T<20K$ may invalidate this result.

**4. Conclusions**

We have measured the heat capacity of isotope enriched Si samples $^{28}$Si, $^{29}$Si and $^{30}$Si as well as Si with the natural isotope composition. The results are compared with recent calculations based on *ab initio* lattice dynamics. The logarithmic derivative of $Cp/T^3$ with respect to the isotope mass has been determined from the experimental data and found to be in good agreement with the calculations above 20K. The maximum occurring below this temperature may be shifted to slightly higher temperatures in the experimental results.

**Acknowledgements**

We thank S. Höhn for experimental assistance and M. Asen-Palmer for a critical reading of the manuscripts. A. Gibin gratefully acknowledges the financial support of the Max-Planck-Society during a stay in Stuttgart and M. Cardona the hospitality of the ICMBA (Barcelona) while performing part of this work and financial support from ICREA (Catalonia).